\documentstyle[amssymb,aps,prb,graphicx]{revtex}

\begin{document}
\title{Charge kinks as Raman scatterers in quarter-filled ladders}
\author{E. Ya. Sherman and C. Ambrosch-Draxl}
\address{Institute for Theoretical Physics, Karl-Franzens-University of Graz, A-8010,%
\\
Graz, Austria}
\author{P. Lemmens and G. G\"{u}ntherodt}
\address{2. Physikalisches Institut, RWTH Aachen, 55099 Aachen, Germany}
\author{P.H.M. van Loosdrecht}
\address{Dept.~of Solid State Physics \& Materials Science Center, \\
University of Groningen, 9747 AG Groningen, the Netherlands}

\maketitle

\begin{abstract}
Charge kinks are considered as fundamental excitations in quarter-filled
charge-ordered ladders. The strength of the coupling of the kinks to the
three-dimensional lattice depends on their energy. The integrated intensity
of Raman scattering by kink-antikink pairs is proportional to $\phi ^{5}$
or $\phi ^{4},$ where $\phi $ is the order parameter. The exponent is
determined by the system parameters and by the strength of the electron-phonon
coupling.
\end{abstract}

An interplay of spin, charge, and lattice degrees of freedom drives a
variety of phase transitions in ladder-like compounds \cite{Rice}. The best
known examples of these systems are half-filled SrCu$_{2}$O$_{3}$ and Sr$%
_{14}$Cu$_{24}$O$_{41}$, where the magnetic properties are determined by $%
s=1/2$ spins at the Cu sites with a superexchange interaction arising from
hopping via oxygen orbitals. Ca-substitution in Sr sites introduces mobile
holes and leads to superconductivity in Sr$_{2.5}$Ca$_{11.5}$Cu$_{24}$O$_{41}
$ under pressure \cite{Nagata}. The pairing interaction has a magnetic
origin since it is favorable to build up a local spin singlet of two holes
if the exchange interaction overcomes the kinetic energy effects \cite
{Dagotto}. In Sr$_{14-x}$Ca$_{x}$Cu$_{24}$O$_{41}$ the dopant holes reside
at the oxygen rather than at the Cu orbitals responsible for magnetic
properties. Another class of ladders are the half-filled vanadates such as
CaV$_{2}$O$_{5}$ and MgV$_{2}$O$_{5}$, where spins 1/2 are located at the V
sites. A more interesting realization is given by NaV$_{2}$O$_{5}$ \cite
{Smolinski} and LiV$_{2}$O$_{5},$ which have VO$_{5}$ pyramids as a common
structure element with the half-filled vanadates. As follows from the
chemical formula, the mean charge of V in NaV$_{2}$O$_{5}$ and LiV$_{2}$O$%
_{5}$ is 4.5. Although similar in stoichiometry and unit cell structure,
these two compounds have strikingly different physical properties. In LiV$%
_{2}$O$_{5},$ the V ions are arranged in parallel chains of non-magnetic V$%
^{5+}$ and magnetic V$^{4+}$. In NaV$_{2}$O$_{5}$ all V ions are equivalent,
making it a rare realization of the quarter-filled ladder \cite
{Smolinski,Horsch}. Only for temperatures below $T_{{\rm co}}\approx 34$ K 
\cite{Isobe,Ohama} the compound undergoes a charge ordering transition into
a spin-gapped phase, which is due to an interplay of charge, spin, and
lattice degrees of freedom \cite{Fagot}. It has been proposed that the
spin-lattice interaction in NaV$_{2}$O$_{5}$ \cite{Sherman} is due to a
strong electron-phonon coupling \cite{Polbianc,Sandvick}.

Low-energy excitations in the ladders related to charge and spin degrees of
freedom and having energies less than 0.2 eV, are well suited for
investigations by light scattering spectroscopy \cite{Fischer,pvl}. However,
the mechanism of their Raman activity, being strongly related to the way
they modulate the crystal's polarizability, requires a detailed
investigation for each excitation.

Since the ladders are one-dimensional systems with strong interactions,
non-linear models appeared to be very useful for investigations of charge-
and lattice- \cite{Krumhansl}, and magnetic \cite{Affleck,Kampf}
excitations. At small doping, when the interaction between the carriers is
not very strong, the generalized $t-J$ model already contains the essential
physics necessary to describe the ladders \cite{Dagotto}. In view of the
large doping in the quarter filled ladders (a depletion of 1/4 hole per
magnetically active site) the excitations in the charge ordered state should
again be considered within a non-linear framework. In this paper, we study
the excitations in a charge ordered quarter-filled ladder system with strong
coupling to the lattice. The elementary excitations are charge kinks
involving ion displacements. We establish a mechanism for Raman scattering
from these excitations and calculate the scattering intensity which is found
to be strongly dependent on the order parameter.

In a quarter-filled ladder each rung $i$ is occupied by a charge $%
Q_{i}=Q_{i}^{\left| L\right\rangle }+Q_{i}^{\left| R\right\rangle }=1,$
where $\left| L\right\rangle $ and $\left| R\right\rangle $ denotes the left
and right leg state, respectively. The electronic degree of freedom is the
hopping of the charge between two ions of a rung described by a matrix
element $t_{\perp }$. Hopping between nearest-neighbor rungs is prohibited
since it requires the energy of approximately $2t_{\perp }$ to put electrons
from two binding two-site states, with the energies of $-t_{\perp }$ each,
to two one-site states at one rung. The wave function of an electron on a
rung has the form $\psi _{i}=\alpha _{L}|L\rangle +\alpha _{R}|R\rangle ,$
where $\alpha _{L}^{2}+\alpha _{R}^{2}=1$. In the disordered phase $\alpha
_{L}=-\alpha _{R}=1/\sqrt{2}$.

The total energy of the system $E_{{\rm tot}}=U_{{\rm cor}}+K+U_{{\rm lat}}$
is the sum of the correlation energy $U_{{\rm cor}}$ due to the repulsion of
electrons at nearest-neighbor sites described by $V_{{\rm cor}}$, the
kinetic energy $K$, and the lattice term $U_{{\rm lat}}$ which includes the
electron-phonon coupling. $U_{{\rm cor}}$ is the sum over the nearest
neighbor interactions of the type $V_{{\rm cor}}\left( Q_{i}^{\left|
L\right\rangle }Q_{i+1}^{\left| L\right\rangle }+Q_{i}^{\left|
R\right\rangle }Q_{i+1}^{\left| R\right\rangle }\right) $. With $\phi
_{i}\equiv \phi _{i}^{L}=Q_{i}^{L}-1/2,$ this contribution can be written
as: $U_{{\rm cor}}=V_{{\rm cor}}\sum_{i}\phi _{i}\phi _{i+1}$, where $\phi
_{i}^{L}=-\phi _{i}^{R}$ has been used. The kinetic energy per rung can be
expressed as: $K_{i}=-t_{\perp }+2t_{\perp }\left( \phi _{i}^{2}+\phi
_{i}^{4}\right) $. The lattice term \cite{Sherman,Polbianc} has the
Holstein-like form $U_{{\rm lat}}=C\sum {\phi _{i}^{l}z_{i,l}}+\kappa \sum {%
z_{i,l}^{2}/2}.$ The elastic contribution to the energy due to a
displacement $z_{i,l}$ of the ions at rung $i$ and leg $l$ is $%
1/2\sum_{i,l}\kappa z_{i,l}^{2}$, with lattice force constant $\kappa
=M\Omega ^{2}$, where $M$ is the mass of the atom, $\Omega $ the vibrational
frequency, and $C$ is the deformation potential. The static ion displacement
minimizing $U_{{\rm lat}}$ is ${z_{i}=-}\left( C/\kappa \right) {\phi _{i}}$.

The equilibrium state of the system minimizes $E_{{\rm tot}}$. The result is
a zigzag ordered state \cite{Seno,Mostovoy,Thalmeier} with the order
parameter $\phi _{i}=\phi _{{\rm lat}}\cdot (-1)^{i},$ where $\phi _{{\rm lat%
}}^{2}=\left( 2(V_{{\rm cor}}-t_{\perp })+V_{{\rm lat}}\right) /4t_{\perp },$
and $V_{{\rm lat}}=C^{2}/\kappa .$ The subscript ''{\rm lat }'' serves as a
reminder that the lattice is involved in the charge ordering. The phase
transition occurs when $2V_{{\rm cor}}+V_{{\rm lat}}>2t_{\perp }$.

In order to study the dynamics, the energy $E_{{\rm tot}}$ is rewritten in a
continuum approximation, with a time-dependent order parameter $\phi (y,t)$
that slowly changes along the ladders. The energy acquires a form typical
for the standard $\phi ^{4}$ field \cite{Rajaraman}: 
\begin{equation}
E\left[ \phi \right] =V_{{\rm cor}}a_{\Vert }^{2}\cdot \int_{-\infty
}^{\infty }\left[ \left( \partial \phi /s_{{\rm lat}}\partial t\right)
^{2}+\left( \partial \phi /\partial y\right) ^{2}+\lambda \left( \phi
^{2}-\phi _{{\rm lat}}^{2}\right) ^{2}/2\right] \frac{dy}{a_{\Vert }}, 
\nonumber
\end{equation}
where $1/s_{{\rm lat}}^{2}=MC^{2}/V_{{\rm cor}}a_{\Vert }^{2}\kappa ^{2}$
with $a_{\Vert }$ being the lattice constant along the ladder, and $\lambda
=4t_{\perp }/V_{{\rm cor}}a_{\Vert }^{2}$. The $\partial \phi /\partial t$
term in Eq.(1) arises due to the kinetic energy of the V ions which
displacements follow their charges since $\partial z_{i}/\partial t=-\left(
C/\kappa \right) \partial {\phi }_{i}/{\partial t.}$ The ground state for $E%
\left[ \phi \right] $ is $\phi (y)\equiv \pm \phi _{{\rm lat}}$. The
classical excitations within this model are kinks and antikinks of the form: 
\begin{equation}
\phi (y,t)=\pm \phi _{{\rm lat}}\tanh {\phi _{{\rm lat}}}\sqrt{\frac{\lambda 
}{2}}\cdot \frac{y-ut}{\sqrt{1-u^{2}/s_{{\rm lat}}^{2}}}.
\end{equation}
Here $u$ is the kink velocity, and its energy $E_{u}=\gamma _{{\rm lat}}E_{%
{\rm lat}}$, where $E_{{\rm lat}}=4\sqrt{2}\sqrt{t_{\perp }V_{{\rm cor}}}%
\phi _{{\rm lat}}^{3}/3,$ and $\gamma _{{\rm lat}}=1/\sqrt{1-u^{2}/s_{{\rm %
lat}}^{2}}$. The density of states of the one-kink excitation is given by: $%
\nu (E)=E/\pi s_{{\rm lat}}\sqrt{E^{2}-E_{{\rm lat}}^{2}}$. The
displacements of ions in the ground state and in a kink excitation are shown
in Fig. 1.

The above consideration is valid for a ``soft'' lattice, where ion
displacements follow the electron redistribution during the kink
propagation. The ``soft'' lattice condition is $\tau _{{\rm k}}\Omega \gg 1,$
where the time $\tau _{{\rm k}}$ characterizes the rate of the change of the
order parameter. It is determined by the kink width $w_{{\rm k}}\sim
1/\gamma _{{\rm lat}}\phi _{{\rm lat}}\sqrt{\lambda }$ as $\tau _{{\rm k}%
}\sim w_{{\rm k}}/u.$ In the ``ultrarelativistic'' limit, where $%
u\rightarrow s_{{\rm lat}},$ the condition $\tau _{{\rm k}}\Omega \gg 1$
cannot be fulfilled and the lattice becomes ``rigid'' due to the Lorentz
contraction of the kink width $\sim 1/\gamma _{{\rm lat}}$.

To understand the difference between a soft and a rigid lattice, let us
consider the condition $\tau _{{\rm k}}\Omega \gg 1$ in more detail. Let us
assume first that the ``light'' velocity $s_{{\rm lat}}$ is large enough so
that $\Omega /\phi _{{\rm lat}}\sqrt{\lambda }s_{{\rm lat}}\ll 1$. In this
case the limiting velocity $u_{\max },$ determined by the condition $\tau _{%
{\rm k}}\sim \Omega ^{-1},$ is $u_{{\rm max}}\sim \Omega /\phi _{{\rm lat}}%
\sqrt{\lambda }\ll s_{{\rm lat}}$. At $u>u_{{\rm max}}$ the kinks become
decoupled from the lattice and propagate on the background of ions displaced
as in the equilibrium charge ordered state. The order parameter for the
rigid lattice $\phi _{{\rm el}}^{2}=(V_{{\rm cor}}-t_{\perp })/2t_{\perp }$
is determined by the electronic subsystem only, and the ``light'' velocity $%
s_{{\rm el}}\gg s_{{\rm lat}}$ since the ions are not involved in the kink
motion anymore. It seems that the excitation energy drops down to
approximately $E_{{\rm el}}=E_{{\rm lat}}\phi _{{\rm el}}^{3}/\phi _{{\rm lat%
}}^{3}$ and becomes weakly $u-$dependent. However, due to the increase of
the ``light'' velocity, the Lorentz contraction disappears and the soft
lattice regime is restored. This fact implies that the lattice becomes rigid
only for the fast kinks with $u$ close to $\Omega /\phi _{{\rm el}}\sqrt{%
\lambda }$. In other words, there are no kinks in the interval $\Omega /\phi
_{{\rm lat}}\sqrt{\lambda }\lesssim u\lesssim \Omega /\phi _{{\rm el}}\sqrt{%
\lambda }$.

In the opposite case when $\Omega /\phi _{{\rm lat}}\sqrt{\lambda }s_{{\rm %
lat}}\gg 1,$ the lattice becomes rigid when $u_{\max }$ is very close to $s_{%
{\rm lat}},$ such that $s_{{\rm lat}}-u_{\max }<$ $s_{{\rm lat}}\left(
\Omega /\phi _{{\rm lat}}\sqrt{\lambda }s_{{\rm lat}}\right) ^{-2}$.
Therefore, the kinks involving lattice displacements are well defined up to
high energies $E_{u_{\max }}\sim E_{{\rm lat}}\Omega /\phi _{{\rm lat}}\sqrt{%
\lambda }s_{{\rm lat}}$. Here the gap in the allowed $u$ is determined by $%
s_{{\rm lat}}<u<\Omega /\phi _{{\rm el}}\sqrt{\lambda }$. If $V_{{\rm cor}%
}<t_{\perp },$ the kink-like electronic excitations are not well defined
being strongly overdamped, and the ``rigid lattice'' electronic part of the
spectrum is absent.

Now consider light scattering by the kinks. These excitations modulate the
charge density and the crystal's dielectric function $\epsilon _{\omega
}^{\beta \eta },$ where $\omega $ is the light frequency, and $\beta $ and $%
\eta $ are Cartesian indices, thereby causing inelastic light scattering at
frequencies equal to the excitation energy.

The variation of the dielectric function is proportional to the square of
the order parameter, and can be written for one rung as (the Cartesian
indices are omitted): 
\begin{equation}
\epsilon _{\omega }\left( \phi _{i}^{2}\right) -\epsilon _{\omega }\left(
\phi _{{\rm lat}}^{2}\right) =\frac{\partial \epsilon _{\omega }}{\partial
\phi ^{2}}\left( \phi _{i}^{2}-\phi _{{\rm lat}}^{2}\right) .
\end{equation}
The change of the polarizability per kink $\phi (y,t)$ from Eq.(2) is: 
\begin{equation}
\Delta \epsilon _{{\rm kink}}=\frac{\partial \epsilon _{\omega }}{\partial
\phi ^{2}}\int \left[ \phi ^{2}(y,t)-\phi _{{\rm lat}}^{2}\right] \frac{dy}{%
a_{\Vert }}=-2\sqrt{2}\frac{\partial \epsilon _{\omega }}{\partial \phi ^{2}}%
\frac{\phi _{{\rm lat}}}{\gamma _{{\rm lat}}\sqrt{\lambda }}.
\end{equation}
With increase of the energy, $\Delta \epsilon _{{\rm kink}}$ decreases due
to Lorentz contraction of the kink width.

Because of the very small photon wavevector, Raman scattering probes
excitations with zero net momentum. Therefore, the Raman active
quasiparticles are kink-antikink pairs with velocities $u$ and $-u,$
respectively. The corresponding contribution of the kink-antikink pairs to
the polarizability $\epsilon _{\omega }^{\beta \eta }$ is shown in Fig. 2
for two different kink energies. The measured spectral density of the
scattered light $\rho (E)$ as a function of energy transfer to the system
(Raman shift) is proportional to the probability of the excitation of the
pair with the total energy $E=2E_{u}$. Since the Raman scattering is a
process of the decay of the incident photon into a continuum consisting of
the scattered photons and electronic excitations, its probability is given
by Fermi's Golden Rule as: 
\begin{equation}
\rho _{{\rm kink}}(2E)=2\pi \nu (E)\left( \Delta \epsilon _{{\rm kink}%
}\right) ^{2}=16\left( \frac{\partial \epsilon _{\omega }}{\partial \phi ^{2}%
}\right) ^{2}\frac{\phi _{{\rm lat}}^{2}}{\lambda s_{{\rm lat}}\gamma _{{\rm %
lat}}\sqrt{\gamma _{{\rm lat}}^{2}-1}}.
\end{equation}
The spectral density of the scattered light in Eq.(5) has a threshold at $%
2E_{{\rm lat}}$ \cite{Rostiashvili} and decreases at $E\gg E_{{\rm lat}}$ as 
$E^{-2}$. The integrated intensity of the Raman continuum of kinks in the
``soft'' lattice can be written as: 
\begin{equation}
I_{{\rm kink}}=\int_{E_{{\rm lat}}}^{\infty }\rho (2E)dE=8\pi \left( \frac{%
\partial \epsilon _{\omega }}{\partial \phi ^{2}}\right) ^{2}\frac{\phi _{%
{\rm lat}}^{2}}{\lambda s_{{\rm lat}}}E_{{\rm lat}}.
\end{equation}
The integration in Eq.(6) was extended to infinity leading to $I_{{\rm kink}%
}\sim $ $\phi _{{\rm lat}}^{2}E_{{\rm lat}}\sim \phi _{{\rm lat}}^{5}$. \
This can be done if $\Omega /\phi _{{\rm lat}}\sqrt{\lambda }s_{{\rm lat}%
}\gg 1,$ which is, as we will see below, the case for NaV$_{2}$O$_{5}$. In
the opposite case of $\Omega /\phi _{{\rm lat}}\sqrt{\lambda }s_{{\rm lat}%
}\ll 1,$ the integration should be performed up to $E\left( \Omega /\phi _{%
{\rm lat}}\sqrt{\lambda }\right) ,$ which is close to $E_{{\rm lat}}$ that
yields $I_{{\rm kink}}\sim $ $\phi _{{\rm lat}}^{4}.$

Let us now discuss a contribution of the kink-antikink excitations to the
Raman spectra of NaV$_{2}$O$_{5},$ which contains a continuum \cite
{Fischer,pvl} and several phonon peaks. The changes at $T<T_{{\rm co}}$
manifest themselves by changes in the phonon frequencies and line shapes,
and by the appearance of new intense peaks. Some of these peaks are
vibrational modes while others can be attributed to magnetic excitations 
\cite{Fischer}. The modification of the continuum consists of two effects: a
depletion at the spectral range up to 30 meV and a moderate redistribution
of the spectral weight \ at higher energies. This observation implies that
the Raman scattering mechanism in the ordered and disordered phases must be
virtually the same, despite the different character of the excitations.

The relevant part of the unit cell of the quarter filled ladder compound NaV$%
_{2}$O$_{5}$ is shown in Fig.3. The Holstein-like electron-phonon coupling
is due to the O3 ions located either above or below the V-O-V rungs \cite
{Sherman} with the hopping matrix element $t_{\perp }\approx 0.35$ eV.
Coupling to the lattice favors the charge ordering since $V_{{\rm lat}}>0$,
however it does not drive the transition itself since in the absence of
correlations the lattice is stable, that is $V_{{\rm lat}}<2t_{\perp }$ \cite
{Sherman}. \bigskip\ 

The charge ordering parameter $\phi _{{\rm lat}}$ determined from the
magnetic susceptibility is approximately $0.2$ \cite{Gros}. Another
possibility to find $\phi _{{\rm lat}}$ is provided by structural data \cite
{Luedecke} where one finds $\left| {z_{i}}\right| {\approx 0.04}$ \AA .
Assuming $C=5$ eV/\AA , and $\Omega =400$ cm$^{-1}$ one obtains $\kappa
\approx 40$ eV/\AA $^{2}$, $V_{{\rm lat}}\approx 0.6$ eV, and ${\phi }_{{\rm %
lat}}=\left( \kappa {/C}\right) {z_{i}\approx 0.3}$. We accept this
magnitude of ${\phi }_{{\rm lat}}$ for further estimates and for $t_{\perp
}=0.35$ eV, and $V_{{\rm cor}}=t_{\perp }$ obtain $E_{{\rm lat}}\approx 20$
meV. \ Therefore, one can expect a depletion in the scattering intensity for
the Raman shift less than $2E_{{\rm lat}}=40$ meV, in agreement with the
experimental data \cite{Fischer}. The chosen model parameters and $a_{\Vert
}=4$\AA\ yield $\Omega /\phi _{{\rm lat}}\sqrt{\lambda }s_{{\rm lat}}\approx
10$, that is the lattice is soft up to high kink energies. This fact
justifies an extension of the integration in Eq.(6) to infinity. Considering
the kinks, we neglected the interaction between different ladders. As we
established above, at $V_{{\rm cor}}<t_{\perp },$ the ``rigid lattice'' part
of the kink spectrum is absent. We assume that this is the case in NaV$_{2}$O%
$_{5}$ since $V_{{\rm cor}}+V_{{\rm lat}}/2-t_{\perp }$ should be
considerably smaller than $t_{\perp }$ to ensure a small order parameter.
For this reason we did not consider scattering by the high-energy kinks
uncoupled to the lattice.

To have a reference point, we compare the spectral densities of the two-kink
scattering and two other Raman processes relevant for this compound. The
first one is the first-order phonon Raman scattering. The estimate of
spectral density at the phonon peak maximum is given by: $\rho _{{\rm ph}%
}\sim \left( \partial \epsilon _{\omega }/\partial {\cal Q}\right)
^{2}\left( z_{0}/a\right) ^{2}a_{\parallel }/\Gamma ,$ where ${\cal Q}$ is
the ratio of the ion displacement to the characteristic lattice constant $a$%
, $z_{0}$ is the zero-point vibrational amplitude for the mode, and $\Gamma $
is the phonon linewidth. The other mechanism is two-magnon Raman scattering
due to the frustrated superexchange interaction. Although the scattering
intensity depends on many details \cite{Muthukumar}, an estimation within
the Fleury-Loudon theory \cite{Fleury} can be done as: $\rho _{{\rm magn}%
}\sim (J/E_{{\rm 0}})^{2}a_{\parallel }/J,$ where $J$ is the superexchange
and $E_{{\rm 0}}$ is of the order of magnitude of the interband transition
energy. By taking into account that $s_{{\rm lat}}a_{\Vert }^{-1}\sim \Omega
,$ the  ratio of intensities can be estimated as: 
\begin{equation}
\rho _{{\rm kink}}:\rho _{{\rm ph}}:\rho _{{\rm magn}}\sim \left( \frac{%
\partial \epsilon _{\omega }}{\partial \phi ^{2}}\right) ^{2}\frac{{\phi }_{%
{\rm lat}}^{2}}{\Omega }:\left( \frac{\partial \epsilon _{\omega }}{\partial 
{\cal Q}}\right) ^{2}\left( \frac{z_{0}}{a}\right) ^{2}\frac{1}{\Gamma }%
:\left( \frac{J}{E_{{\rm 0}}}\right) ^{2}\frac{1}{J}.
\end{equation}
An reasonable value for $\partial \epsilon _{\omega }/\partial \phi ^{2}$ is
about $0.1$, that is for the fully ordered state the change in the
dielectric function would be about 0.1. This is consistent with the
ellipsometric data on NaV$_{2}$O$_{5}$ \cite{Presura,Rubhausen}. \ Eq.(7)
does not contain an evident small parameter which could allow to
discriminate reliably intensities of the different contributions. Dominating
phonon peaks arise due to their small damping $\Gamma \ll \Omega \sim J$
rather than due to high efficiency of the phonon Raman scattering. For
realistic parameters $\partial \epsilon _{\omega }/\partial {\cal Q}=1,$ $%
\left( z_{0}/a\right) ^{2}=10^{-4},$ and $\Gamma =2$ meV, $\rho _{{\rm ph}}$
at the phonon peak is of the same order of magnitude as $\rho _{{\rm kink}}$
at $E\sim 2E_{{\rm lat}}.$ This corresponds well to the experimental data 
\cite{Fischer}, especially for the light polarized along the rungs, assuming
that the continuum in NaV$_{2}$O$_{5}$ is formed by the kink-antikink
excitations. \ \ The two-magnon mechanism is responsible for the Raman
background in LiV$_{2}$O$_{5},$ where the kink-antikink pairs cannot be
excited due to long-range order of nonequivalent V$^{4+}$ and V$^{5+}$ ions.
In this compound the continuum is weak compared to phonons \cite{Popovic},
however, their relative spectral densities are of the same order of
magnitude as in NaV$_{2}$O$_{5}$ when the incident and scattered light are
polarized along the ladders.

At $T>T_{{\rm co}}$ the low-energy electronic excitations are long-ranged
overdamped fluctuations of the order parameter coupled to the dynamical
lattice distortion. These excitations have the same mechanism of Raman
scattering as the kink-antikink pairs considered above, that is modulation
of the polarizability due to intra-rung charge fluctuations leading to a
broad intense continuum. The fluctuations of the order parameter in the form
of dynamical charge ordering persist up to $T\sim 80$K, as it was observed
in Raman spectroscopy \cite{Fischer}, structural X-ray scattering \cite{Ravy}
and spin resonance \cite{Nojiri} experiments. However, at present, it is
hard to separate the contributions of the charge fluctuations and the
two-magnon scattering to the Raman continuum.

In conclusion, we investigated the low energy kink dynamics and Raman
scattering by kink-antikink excitations in the charge ordered phase of a
quarter-filled ladder with strong coupling to the lattice and considered NaV$%
_{2}$O$_{5}$ as an example. The spectral density of the scattered light and
the overall scattering intensity by kink-antikink pairs very strongly
depends on the charge order parameter. The mechanism corresponds well to
three observed features of the continuum: (i) spectral range, (ii) depletion
at Raman shifts less then 30 meV, and (iii) relative intensity to phonons.
However, it is not possible at present to distinguish contributions of the
kink-antikink and the conventional two-magnon processes.

This work is supported by the DFG through SFB341. E.Ya.S. acknowledges
support from the Austrian Science Fund via the project M534-TPH and M591-TPH
and valuable discussions with P. Monod and S. Ravy. P.H.M.v.L. acknowledges
support by INTAS (99-0155).

\begin{figure}
\centering\includegraphics[width=10cm]{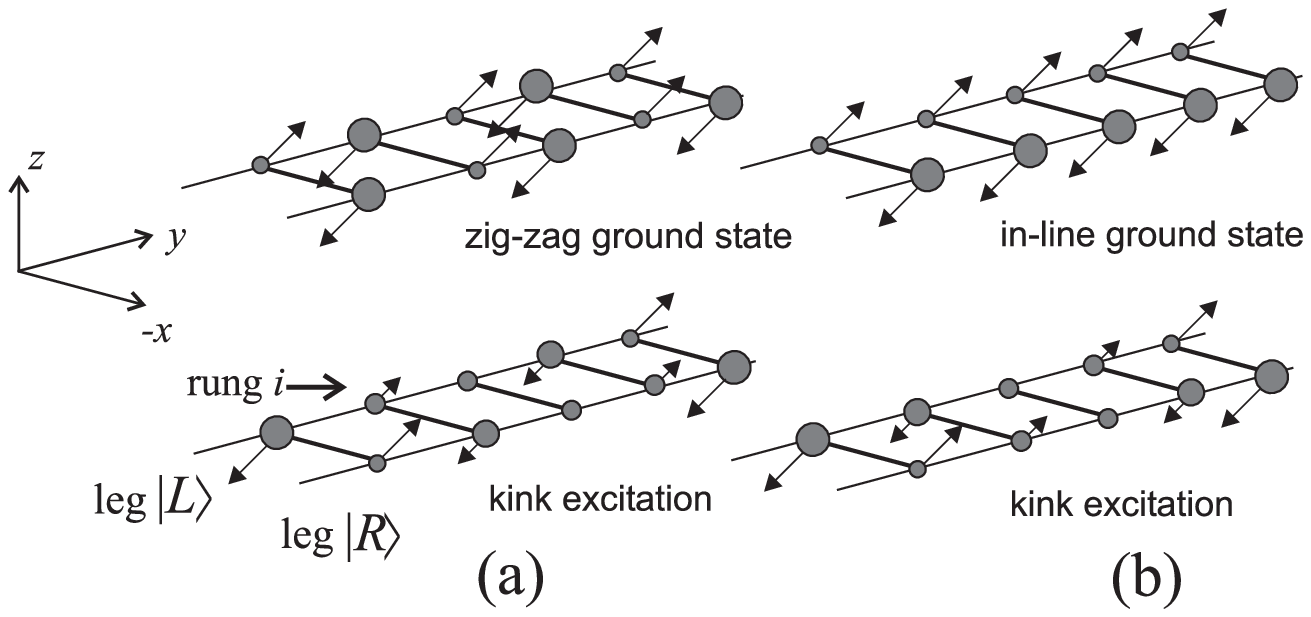}
\caption{ (a) Displacements of V ions (proportional to the length of
arrow) in the zig-zag ground state and in a kink-like excitation. The radii
of the grey circles correspond to the charges of the V ions. (b) The same
for the in-line ordering, presented for comparison to (a).}
\end{figure}

\vspace{3cm}

\begin{figure}
\centering\includegraphics[width=10cm]{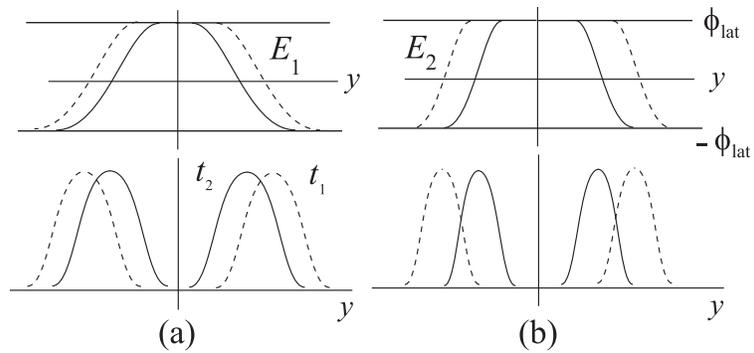}
\caption{The order parameter $\phi (y,t)$ in kink-antikink pairs for
different kink energies (upper panel) and absolute values of the changes in
the dielectric function (Eq.(3)) caused by the pairs (lower panel) for time $%
t_{1}$ (solid line) and $t_{2}>t_{1}$ (dashed line). (a) Kink energy $E_{1},$
(b) kink energy $E_{2}>E_{1}.$}
\end{figure}

\clearpage

\begin{figure}
\centering\includegraphics[width=10cm]{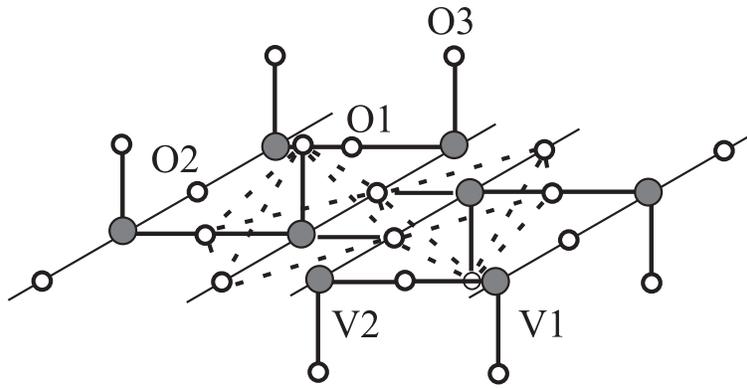}
\caption{
Schematic plot of a part of the crystal structure of NaV$_{2}$O$_{5}$. 
Na ions are not shown in the Figure. The Holstein-like interaction arises
due to the asymmetry of the unit cell related to the position of the O3 ion.
}
\end{figure}

\end{document}